\newcommand{\be}[0]{\begin{equation}}
\newcommand{\ee}[0]{\end{equation}}
\newcommand{\ba}[0]{\begin{eqnarray}}
\newcommand{\ea}[0]{\end{eqnarray}}

\newcommand\MeV{\,\mbox{MeV}}

\def\ie{{\it i.e.}}

\documentclass[report]{JHEP3}

\usepackage{epsfig,multicol}
\newcommand\fverb{\setbox\pippobox=\hbox\bgroup\verb}
\newcommand\fverbdo{\egroup\medskip\noindent%
            \fbox{\unhbox\pippobox}\ }
\newcommand\fverbit{\egroup\item[\fbox{\unhbox\pippobox}]}
\newbox\pippobox

\title{The NNLO non-singlet QCD analysis of parton distributions based
 on Bernstein polynomials}

\author{Ali N. Khorramian$^{a,c}$ and S. Atashbar
Tehrani$^{b,c}$\\
\llap{$^a$}Physics Department, Semnan University, Semnan, Iran\\
\llap{$^b$}Physics Department, Persian Gulf University, Boushehr,
Iran \\
\llap{$^c$}Institute for Studies in Theoretical Physics and Mathematics (IPM),\\
P.O.Box 19395-5531, Tehran, Iran\\

E-mail: \email{Khorramian@theory.ipm.ac.ir},
\email{atashbar@ipm.ir}}

\abstract{A non-singlet QCD analysis of the structure function
$xF_3$ up to NNLO is performed based on the Bernstein polynomials
approach. We use recently calculated NNLO anomalous dimension
coefficients for the moments of the $xF_3$ structure function in
$\nu N$ scattering. In the fitting procedure, Bernstein polynomial
method is used to construct experimental moments from the $xF_3$
data of the CCFR collaboration in the region of $x$ which is
inaccessible experimentally. We also consider Bernstein averages to
obtain some unknown parameters which exist in the valence quark
densities in a wide range of $x$ and $Q^2$. The results of valence
quark distributions up to NNLO are in good agreement with the
available theoretical models. In the analysis we determined the
QCD-scale $\Lambda^ {\overline{MS}} _{QCD, N_{f}=4}=211$ MeV (LO),
$259$ MeV (NLO) and $230$ MeV (NNLO), corresponding to
$\alpha_s(M_Z^2)=0.1291$ LO, $\alpha_s(M_Z^2)=0.1150$ NLO and
$\alpha_s(M_Z^2)=0.1142$ NNLO. We compare our results for the QCD
scale and the $\alpha_s(M_Z^2)$ with those obtained from deep
inelastic scattering processes. }
\begin{document}
\section{Introduction}\label{section1}

The global parton analysis of deep inelastic scattering (DIS) and
the related hard scattering data are generally performed at next-to
leading order (NLO). Presently the next-to leading order is the
standard approximation for most of the important processes in QCD.
Analyzing DIS at next-to-next-to-leading order (NNLO) is important
as we may be able to investigate the hierarchy LO $\rightarrow$ NLO
$\rightarrow$ NNLO for the processes using the most precise
available data.

The corresponding one- and two-loop splitting functions have been
known for a long time [1-11]. The NNLO corrections should be
included in order to arrive at quantitatively reliable predictions
for hard processes occurring at present and future high-energy
colliders. These corrections are so far known only for the structure
functions in the deep-inelastic scattering [12-15], for  the
Drell-Yan lepton-pair and gauge-boson production in
proton--(anti-)proton collisions [16-19], and the related cross
sections for Higgs production in the heavy-top-quark approximation
[17,20-22].\\

Recently much effort has been invested in computing NNLO QCD
corrections to a wide variety of partonic processes
 and therefore it is needed to generate parton
distributions also at NNLO, so that the theory can be applied in a
consistent manner.  Analysis on the NNLO cross sections for jet
production is under way and it is expected to yield results in the
near future, see Ref.~\cite{Glover:2002gz} and references therein.
For the corresponding three-loop splitting functions, on the other
hand, only partial results have been obtained up to now, most
notably on the lowest six/seven (even or odd) integer-$N$ Mellin
moments [24-26].\\
S. Moch {\it{et al}}. \cite{Moch:2004pa} computed the higher order
contributions up to three-loop splitting functions governing the
evolution of unpolarized non-singlet quark densities in the
perturbative QCD.

During the recent years the interest to use CCFR data
\cite{CCFR:1997} for $xF_3$ structure function in the higher orders,
based on the orthogonal polynomial expansion method has increased
[29-34].

 In this paper we determine the flavor non-singlet
parton distribution functions, $xu_v(x,Q^2)$ and $xd_v(x,Q^2)$,
using the Bernstein polynomial approach up to the NNLO level. This
calculation is possible now, as the non-singlet anomalous dimension
coefficients in $N$-Moment space in three loops has already been
introduced \cite{Moch:2004pa,Vogt:2004ns}.

The plan of the paper is to give an introduction to the CCFR data in
Section 2. In Section 3 we present a brief review of QCD formalism
of the non-singlet structure function in three loops.
Parametrization of parton densities are written down in Section 4.
Section 5 contains a description of the Bernstein polynomial
averages to be employed in the fits. Non-singlet quark distributions
in the $x$-space are illustrated in Section~6. Section 7 contains a
discussion and conclusions.
\\

\section{CCFR experimental data}

The measurements of the CCFR collaboration provide a precise
determination of the non-singlet deep inelastic scattering structure
functions of neutrinos and anti-neutrinos on nucleons,
$xF_{3}(x,Q^2)$. Data for $xF_{3}$ in neutrino-nucleon scattering is
available from the CCFR collaboration \cite{CCFR:1997}. The data was
obtained from the scattering of neutrinos off iron nuclei and the
measurements span the  ranges $1.26\leq
Q^{2}\leq199.5\;{\rm{GeV}}^2$ and $0.015\leq x\leq0.75$.

The $Q^2$-dependence of moments of structure functions can be
predicted in perturbative QCD, and fits to data can be used to infer
$\alpha_s(M_Z^2)$. A difficulty is that there are upper and lower
limits on the experimentally accessible range of $x$ at low and high
$Q^2$, respectively. This is shown in Fig.1 in which we plot the
CCFR data for different values of $Q^2$. We can see that at the
lower range of $Q^{2}$ we are limited to low-$x$ data, and at high
$Q^{2}$ we are limited to the high $x$ range.

\begin{figure}[tbh]
\centerline{\includegraphics[width=0.8\textwidth]{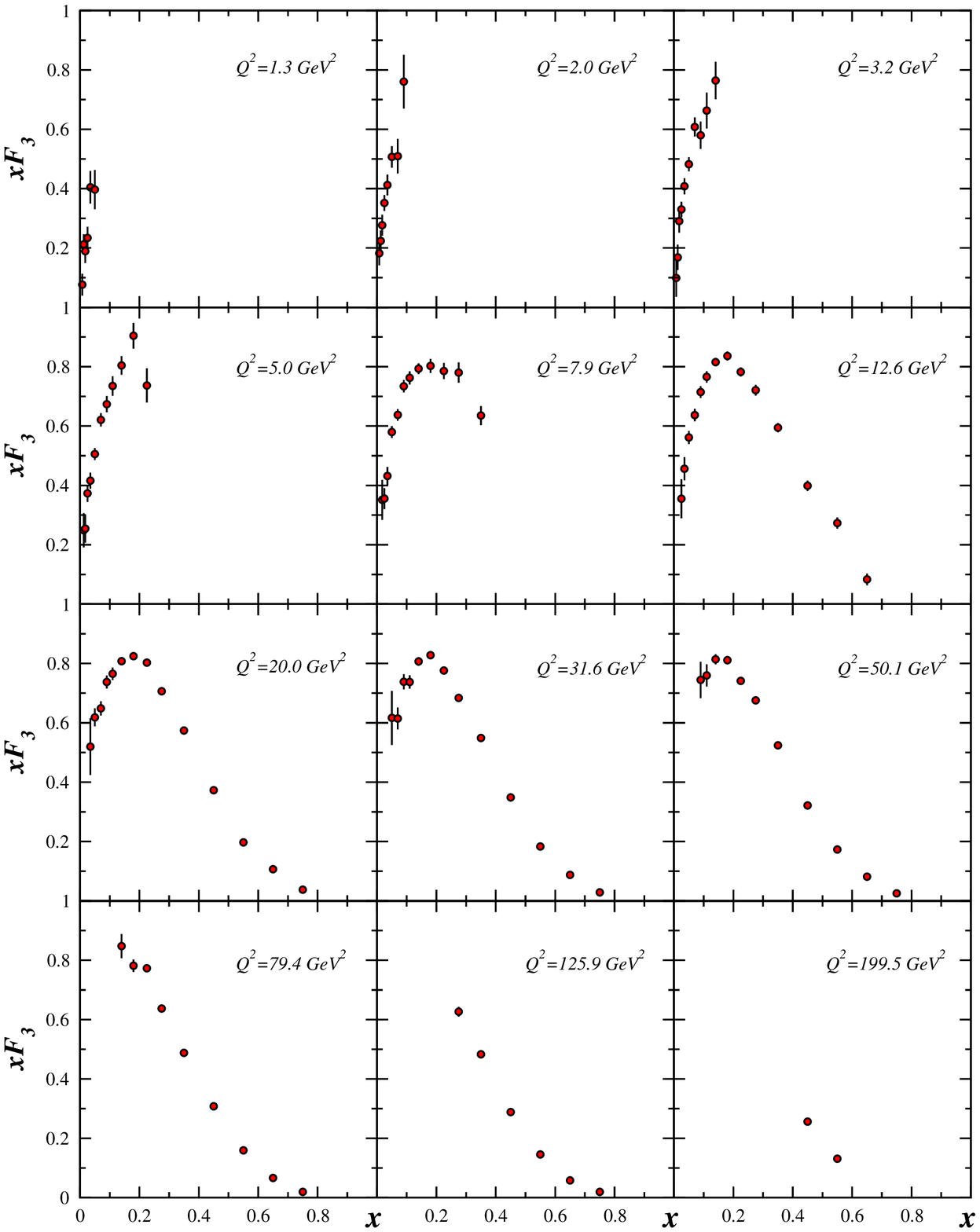}}
\caption{\textsf{CCFR $xF_{3}$ experimental data as a function of
$x$ and for some different values of $Q^{2}$.} }\label{f:CCFRDATA}
\end{figure}
In order to reliably evaluate a moment at a particular $Q^{2}$, we
require data for the whole range of $x$.  In fact for a given
 value of $Q^2$, only  a limited number of experimental points,
 covering a partial range of values of $x$ are available.  The method devised to
 deal with this situation
 is to take  averages of the structure function weighted by suitable
 polynomials.

 Before reconstructing of the structure
 function from moments, we need to know how
the structure function behaves in the missing data region. As we
will see in next sections, in the fitting procedure we need to fit
the $xF_3$ data of the CCFR collaboration.
 In this regard
  we can choose the extrapolation method. In this
method,  we fit $x{F_3}(x,{Q^2})$ for each fixed value of $Q^{2}$
separately to the  phenomenologically convenient expression \be
xF_3^{(phen)}(x)={\cal{A}}x^{\cal{B}}(1-x)^{\cal{C}}\;.
\label{eq:xf3pheno} \ee This form ensures zero values for ${xF_{3}}$
at $x=0$, and $x=1$. The parameters ${\cal{A}}$, ${\cal{B}}$ and
${\cal{C}}$ are obtained by performing $\chi^{2}$ fitting of
Eq.~(\ref{eq:xf3pheno}) to data for $xF_{3}$. They are
$Q^{2}$-dependent quantities, and errors on their values are
obtained by performing the fitting with the data for $xF_{3}$
shifted to the two extremes of the error bars.

 In Table 1 we have presented  the numerical values of
$\cal{A},\cal{B}$ and $\cal{C}$ at $Q^2=20$, $31.6$, $50.1$, $79.4$,
$125.9$ GeV$^2$. We have only included data for ${Q}^{2}{\ge}\;20$
GeV$^2$, this has the merit of simplifying the analysis by avoiding
evolution through flavor thresholds.
\begin{center}
\begin{tabular}{|c|ccc|}
\hline\hline
Q$^2$(GeV$^{2}$) & $\mathcal{A}$ & $\mathcal{B}$ & $\mathcal{C}$ \\
\hline\hline
$20$ & $4.7425 \pm 0.7585$ & $0.6356\pm 0.0638$ & $3.3756 \pm 0.0892$ \\
$31.6$ & $5.4735 \pm 0.8722$ & $0.6936 \pm 0.0456$ & $3.6595 \pm 0.0537$ \\
$50.1$ & $5.6795 \pm 0.4841$ & $0.6977 \pm 0.0843$ & $3.8390 \pm 0.0459$ \\
$79.4$ & $4.5076 \pm 0.5033$ & $0.5666 \pm 0.0808$ & $3.7574 \pm 0.0486$ \\
$125.9$ & $7.0775 \pm 0.8641$ & $0.8186 \pm 0.0475$ & $4.2456 \pm 0.0458 $ \\
\hline\hline
\end{tabular}
{\normalsize \\
 \textsf{\\Table~1: Numerical values of fitting
${\cal{A}},{\cal{B}},{\cal{C}}$ parameters in
Eq.~(\ref{eq:xf3pheno}). \label{tab1fitxf3}}}
\end{center}

\section{QCD formalism}
 To carry out the analogous analysis at
NNLO we need both the relevant splitting functions as well as the
coefficient functions. Now not only the deep inelastic coefficient
functions are known at NNLO, but also the anomalous dimensions in
$N$-Moment space are available at this order
\cite{Moch:2004pa,Vogt:2004ns}.
 In this section we want to
introduce the non-singlet structure function in Mellin moment space
up to three loops order.

The structure function $xF_3$, associated with the parity-violating
weak interaction, represents the momentum density of valence quarks.
So in the LO approximation we can write, \ba
 xF_3^{\nu N}=xu_v(x)+ xd_v(x)+ 2\;xs(x)-2\;xc(x)\;,\\ \nonumber
xF_3^{\bar{\nu} N}=xu_v(x)+ xd_v(x)- 2\;xs(x)+2\;xc(x)\;,
 \ea
where $u_v\equiv u-\bar u$ and $d_v\equiv d-\bar d$ are the proton
valence densities. The asymmetry of the $s-c$ doublet results in $
xF_3^{\nu N}\neq xF_3^{\bar{\nu} N}$. The method of extracting CCFR
experimental data, extracts the average of the neutrino and
anti-neutrino distributions, so that \be
 xF_3(x)=\frac{xF_3^{\nu N}+xF_3^{\bar{\nu} N}}{2}=xu_v(x)+
 xd_v(x)\;,
 \ee
here it is obvious that the  $xF_3(x)$ is related to the combination
of valence quark densities.\\
Let us now define the Mellin moments for the $\nu N$ structure
function $xF_3(x,Q^2)$:
\begin{equation}
\label{eq:momxf3IV} {{\cal{M}}_3^{\nu N}(N,Q^2)}=\int_0^1
x^{N-1}F_3(x,Q^2)dx\;.
\end{equation}
The theoretical expression for these moments obey the following
renormalization group equation \cite{Kataev:1997nc}
\begin{equation}
\label{eq:RGE}
\bigg(\mu\frac{\partial}{\partial\mu}+\beta(A_s)\frac{\partial}
{\partial A_s} +\gamma_{N}^{NS}(A_s)\bigg)
{\cal{M}}(N,Q^2/\mu^2,A_s(\mu^2))=0\;. \label{rg}
\end{equation}
The symbol $A_s$ denotes the strong coupling constant normalized to
$A_s=\alpha_s/(4\pi)$ and is governed by the QCD $\beta$-function as
\begin{eqnarray}
\label{eq:BetaFun} \mu\frac{\partial
A_s}{\partial\mu}=\beta(A_s)=-2\sum_{i\geq 0} \beta_i A_s^{i+2}\;.
\end{eqnarray}
Eq.(\ref{eq:BetaFun}) is solved in the ${\overline{MS}}$-scheme
applying the matching of flavor thresholds at $Q^2=m_c^2$ and
$Q^2=m_b^2$ with $m_c=1.5$ GeV and $m_b=4.5$ GeV as described in
\cite{Chetyrkin:1997sg,Bethke:2000ai}. ${\overline{MS}}$-scheme
convention introduced in \cite{Bardeen:1978yd} is extended in this
way. In order to be able to make a comparison with the other
measurements of $\Lambda_{QCD}$ we adopt this prescription. The
solution of Eq.(\ref{eq:BetaFun}) in the NNLO is given by

\begin{eqnarray}
\label{eq:alfaNNLO}
A_s &=&\frac{1}{\beta _{0}\ln Q^{2}/\Lambda _{\overline{MS}%
}^{2}}-\frac{\beta _{1}\ln (\ln Q^{2}/\Lambda
_{\overline{MS}}^{2})}{\beta
_{0}^{3}(\ln Q^{2}/\Lambda _{\overline{MS}}^{2})^{2}}+  \nonumber \\
&&\frac{1}{\beta _{0}^{5}(\ln Q^{2}/\Lambda
_{\overline{MS}}^{2})^{3}}[\beta _{1}^{2}\ln ^{2}(\ln Q^{2}/\Lambda
_{\overline{MS}}^{2})-\beta _{1}^{2}\ln (\ln Q^{2}/\Lambda
_{\overline{MS}}^{2})+\beta _{2}\beta _{0}-\beta _{1}^{2}]\;.\nonumber \\
\end{eqnarray}Notice that in the above the numerical expressions for $\beta_0$,
$\beta_1$ and $\beta_2$ are
\begin{eqnarray}
\label{eq:beta}
\beta_0&=&11-0.6667f \;, \nonumber \\
\beta_1&=&102-12.6667f \;,\nonumber \\
\beta_2&=&1428.50-279.611f+6.01852f^2\;,
\end{eqnarray}
where $f$ denotes the number of active flavors.\\

 In
Mellin-$N$ space the evolution equation is solved
\cite{Kataev:1999bp}. The non–singlet structure function
${\cal{M}}(N, Q^2)$ is given by \be \label{eq:momxF3NNLO}
{{\cal{M}}(N,Q^2)} = [1+C^{(1)}(N)A_s+C^{(2)}(N)A_s^2] \;
{f^{NS}(N,Q^2)}\;, \ee where the $f^{NS}(N,Q^2)$ is the Mellin
transform of the non-singlet quark combinations and the $C^{(k)}(N)$
are the corresponding Wilson coefficients \cite{Moch:1999eb}. For
the remainder of this paper we simplify our notation by dropping the
sub- and superscript `$\nu N$' and `$(3)$' in Eqs. (\ref{rg},
\ref{eq:momxF3NNLO}).

The solution of the non-singlet evolution for the parton densities
to 3-loop order reads

\begin{eqnarray}
\label{eq:mnsNNLO} {f^{NS}(N,Q^2)}&=&{f^{NS}(N,Q_0^2)}\left(
\frac{A_s(Q^{2})}{A_s(Q_{0}^{2})}\right) ^{\gamma _{0}^{NS}/2\beta
_{0}}\left\{1+[A_s(Q^{2})-A_s(Q_{0}^{2})]\left( \frac{\gamma
_{1}^{}}{2\beta _{1}}-\frac{\gamma _{0}^{NS}}{2\beta _{0}}\right) \frac{%
\beta _{1}}{\beta _{0}} \right. \nonumber \\
&&+[A_s(Q^{2})-A_s(Q_{0}^{2})]^{2}\frac{\beta _{1}^{2}}{8\beta
_{0}^{2}}\left(
\frac{\gamma _{1}^{}}{\beta _{1}}-\frac{\gamma _{0}^{NS}}{\beta _{0}}%
\right) ^{2} \nonumber \\
&&\left. +\frac{1}{4}[A_s^{2}(Q^{2})-A_s^{2}(Q_{0}^{2})]\left( \frac{1}{\beta _{0}}%
\gamma _{2}^{}-\frac{\beta _{1}}{\beta _{0}^{2}}\gamma _{1}^{}+%
\frac{\beta _{1}^{2}-\beta _{2}\beta _{0}}{\beta _{0}^{3}}\gamma
_{0}^{NS}\right) \right \}\;,
\end{eqnarray}
where $f^{NS}$ is the valence quark compositions as
\begin{eqnarray}
\label{eq:VComposi} f^{NS}=(u-\bar u)+(d-\bar d)\;.
\end{eqnarray}
By considering symmetry between sea quark distributions we can write
\begin{eqnarray}
\label{eq:VComposi} f^{NS}(N,Q_0^2)=u_v(N,Q_0^2)+d_v(N,Q_0^2)\;.
\end{eqnarray}

In the next section we will introduce the functional form of the
valence quark distributions and we will parameterize these
distributions at the scale of $Q_0^2$. As we see in Mellin-$N$ space
the non-singlet parts of structure function in the NNLO
approximation, \ie
 $\;{\cal{M}}(N,Q^2)$, can be obtained from
 the corresponding Wilson coefficients $C^{(k)}(N)$  and the
  non-singlet quark
 densities.

\begin{figure}[tbh]
\centerline{\includegraphics[width=0.7\textwidth]{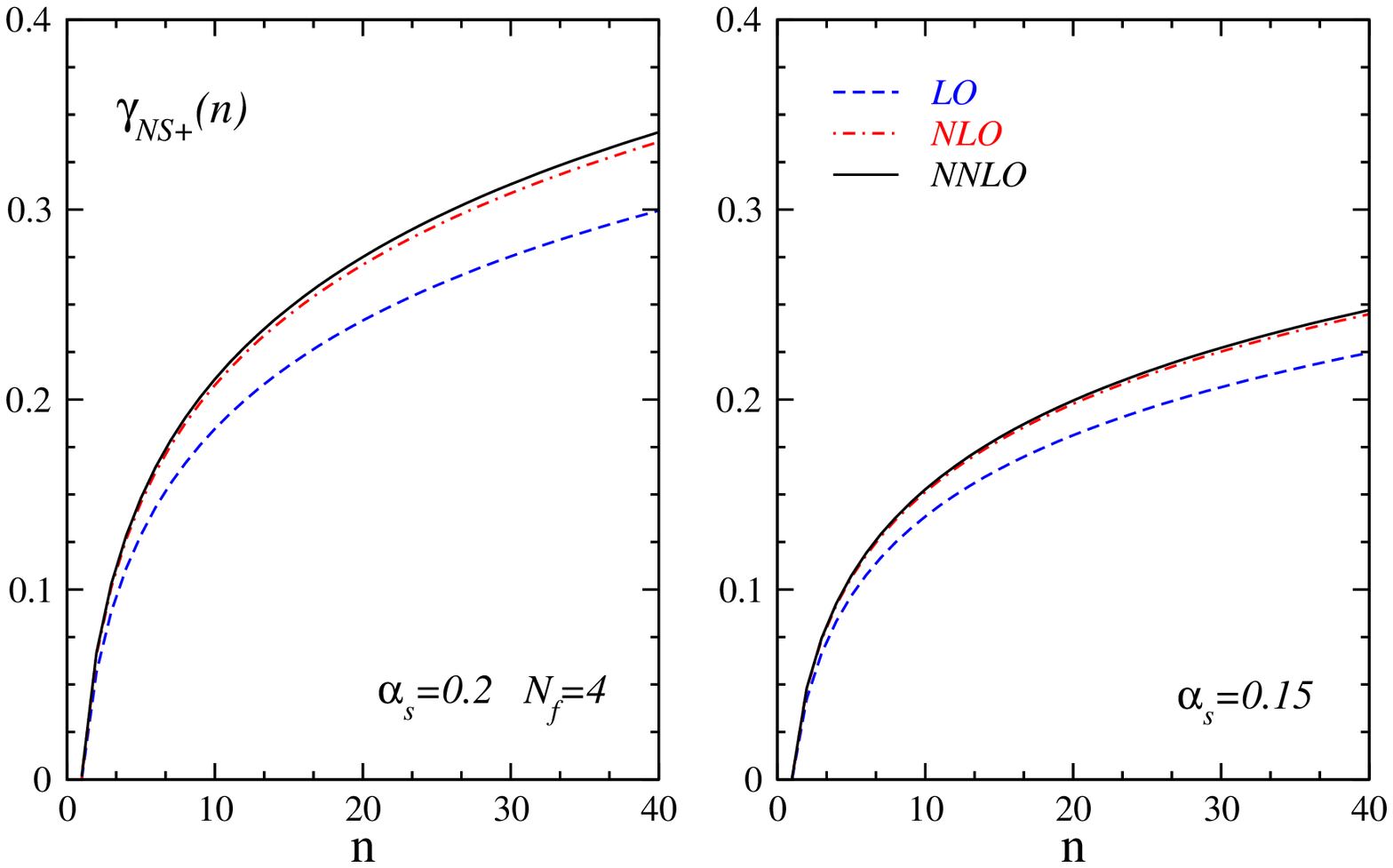}}
\caption{\textsf{The perturbative expansion of the anomalous
dimension
 $\gamma_{_{NS+}}(n)$ for four flavors at $\alpha_s=0.15,0.2$.}}
\label{pic:Fig1-gamma}
\end{figure}

 By using the anomalous dimensions in one, two and three loops
from  \cite{Moch:2004pa} and inserting them in
Eq.~(\ref{eq:mnsNNLO}) and using Eq.~(\ref{eq:momxF3NNLO}), the
moments of non-singlet structure function as a function of $N$ and
$Q^2$ are available. The results of \cite{Moch:2004pa} for
$\gamma_{_{NS+}}(n)$ are depicted in Fig.~\ref{pic:Fig1-gamma} for
four active flavors and typical values $\alpha_s=0.15,0.2$ for the
strong coupling constant.
\vspace{1 cm}

\section{Parametrization of the parton densities}
In this section we discuss how we can determine the valence quark
densities at the input scale of $Q_0^2=1$ GeV$^2$. First of all,
we should notice that the sum of the $u_v$ and $d_v$ distribution
functions can be obtained from the CCFR data and not the two
distributions separately. To separate the $xu_v$ and $xd_v$
contributions to $xF_3$ we need to use the relation between two
distribution functions.

 To start the parameterizations of the
above mentioned parton distributions at the input scale of
$Q_0^2$, we choose the following parametrization for the
$u$-valence quark density

\begin{equation}
\label{eq:xuv}
xu_{v}(x,Q_0^2)=N_{u}x^{a}(1-x)^{b}(1+c\sqrt{x}+dx)\;.
\end{equation}

In the above the  $x^{a}$ term  controls the low-$x$ behavior
parton densities, and the $(1-x)^{b}$ terms the large $x$ values.
The remaining polynomial factor accounts for additional medium-$x$
values. To separate the $xu_v$ and $xd_v$ contributions to $xF_3$
we assume the relation between two distribution functions as
\begin{equation}
\label{eq:ratio}
 \frac{d_v}{u_v}={\cal N}(1-x)^e\;,
\end{equation}
this equation is same as the ratio which has reported in
Ref.\cite{Diemoz:1987xu}. The parameter $e$ in this ratio is very
important to control the behavior of $F_3$ for large $x$ value and
the coefficient ${\cal N}$ is the normalization constant
($=\frac{N_d}{N_u}$). Therefor the parametrization of the
$d$-valence quark density is as follows
\begin{equation}
\label{eq:xdv}
xd_{v}(x,Q_0^2)={\frac{N_d}{N_u}}(1-x)^{e}\;xu_{v}(x,Q_0^2)\;.
\nonumber\end{equation}
Normalization constants $N_u$ and $N_d$ are
fixed by

\begin{equation}
\label{eq:const1} \int_0^1u_v(x)dx=C_u\;,
\end{equation}
\begin{equation}
\label{eq:const2} \int_0^1d_v(x)dx=C_d\;,
\end{equation}
so normalization constants are equal to

\begin{equation}
\label{eq:normuv}
N_u=\frac{C_u}{B(a,1+b)+cB(1/2+a,1+b)+dB(1+a,1+b)}\;,
\end{equation}
\begin{equation}
\label{eq:normdv}
N_d=\frac{C_d}{B(a,1+b+e)+cB(1/2+a,1+b+e)+dB(1+a,1+b+e)}\;,
\end{equation}
\\here $C_u=2$ and $C_d=1$ are respectively the number of $u_v$ and
$d_v$ quarks
and  $B(a,b)=\frac{\Gamma (a)\Gamma (b)}{\Gamma (a+b)}$ is the Euler Beta function.\\
The above normalizations are very effective to control unknown
parameters in Eqs.~(\ref{eq:xuv},\ref{eq:xdv}) via the fitting
procedure. The five parameters with $\Lambda _{QCD}^{N_{f}=4}$ will
be extracted by using the Bernstain polynomials
approach.\\

Using the valence quark distribution functions, the moments of
 $u_{v}(x,Q_0^2)$
and $d_{v}(x,Q_0^2)$ distributions can be easily calculated.  The
Mellin moments for the sum of the two valence quark distributions
in the proton is as follows
\begin{eqnarray}
\label{eq:momxuvQ0} u_{v}(N,Q_0^2)+d_{v}(N,Q_0^2)&=&\int_0^1
x^{N-2}\;\left(xu_{v}(x,Q_0^2)+xd_{v}(x,Q_0^2)\right)\;dx \nonumber \\
&=&\int_0^1
x^{N-2}\;xu_{v}(x,Q_0^2)\left(1+\frac{N_d}{N_u}(1-x)^e\right)\;dx\;.
\end{eqnarray}
Now by inserting the above equation in the
Eq.~(\ref{eq:VComposi}), the function of $f^{NS}(N,Q_0^2)$ is
determined in terms of unknown parameters $a,\;b,\;c,\;d,\;e$.
This function is needed to determine the moments of non-singlet
structure function in the related order.

\section{Reconstruction of the structure function from moments}
 Although it is relatively easy to compute the $N$th moment from the
 structure functions, the inverse process is not obvious. To do
 this, we adopt a mathematically rigorous but easy
 method \cite{Yndu:78} to invert the moments and retrieve the structure
 functions. The method is based on the fact that for a given
 value of $Q^2$, only  a limited number of experimental points,
 covering a partial range of values of $x$ are available.
 The method devised to deal with this situation
 is to take  averages of the structure function weighted by suitable
 polynomials. We define the Bernstein polynomials as follows,
 \be
B_{nk}(x)=\frac{\Gamma (n+2) }{\Gamma (k+1) \Gamma (n-k+1)
}x^k(1-x)^{n-k}\;;\;\;\;\;\;\; n\geq{k}. \ee These polynomials have
a number of useful properties. These functions are normalized such
that $\int_0^1 B_{n,k}(x)dx=1$ and they are also constructed such
that they are zero at endpoints $x=0$ and $x=1$.
  These polynomials
 are positive and have a single maximum located at
\ba
 \bar{x}_{nk}(x)&=&\int_0^1x\;B_{nk}(x)\;dx \nonumber \\ &=&\frac{\Gamma (n+2)\; \Gamma(k+2)}{
\Gamma (n+3)\;\Gamma(k+1)}\;, \ea and finally, they are concentrated
around this point, with a spread of \ba
\Delta{x_{nk}}&&=\left[\int_0^1 (x-\bar{x}_{nk})^2\;B_{nk}(x)\;dx
\right]^{\frac{1}{2}}\nonumber\\\nonumber\\
&&= \sqrt{\frac{\Gamma (n+2)\; \Gamma(k+3)}{ \Gamma
(n+4)\;\Gamma(k+1)}-\left(\frac{{\Gamma (n+2)}\; {\Gamma(k+2)}}{
{\Gamma(n+3)}\;{\Gamma(k+1)}}\right)^2}\;. \ea Therefore, for a
given value of $Q^2$, the Bernstein averages of $F_3$ which are
defined by, \be
F_{nk}(Q^2){\equiv}\int_{0}^{1}dxB_{nk}(x)F_3(x,Q^2)\;,
\label{eq:fnk1} \ee represents an average of the function
$F_3(x,Q^2)$ in the region
${\bar{x}}_{nk}-\frac{1}{2}\Delta{x}_{nk}{\leq}x{\leq}{\bar{x}}_{nk}+
\frac{1}{2}\Delta{x}_{nk}$. The key point is, the values of $F_3$
outside this interval have a small contribution to the above
integral, as $B_{nk}(x)$ tends to zero very quickly. By a suitable
choice of $n$, $k$ we manage to adjust to the region where the
average is peaked around values which
we have experimental data \cite{CCFR:1997}.\\

\begin{figure}
\begin{center}
\begin{tabular}{c c c}
\hspace*{-1truecm}\includegraphics[width=0.3\textwidth]{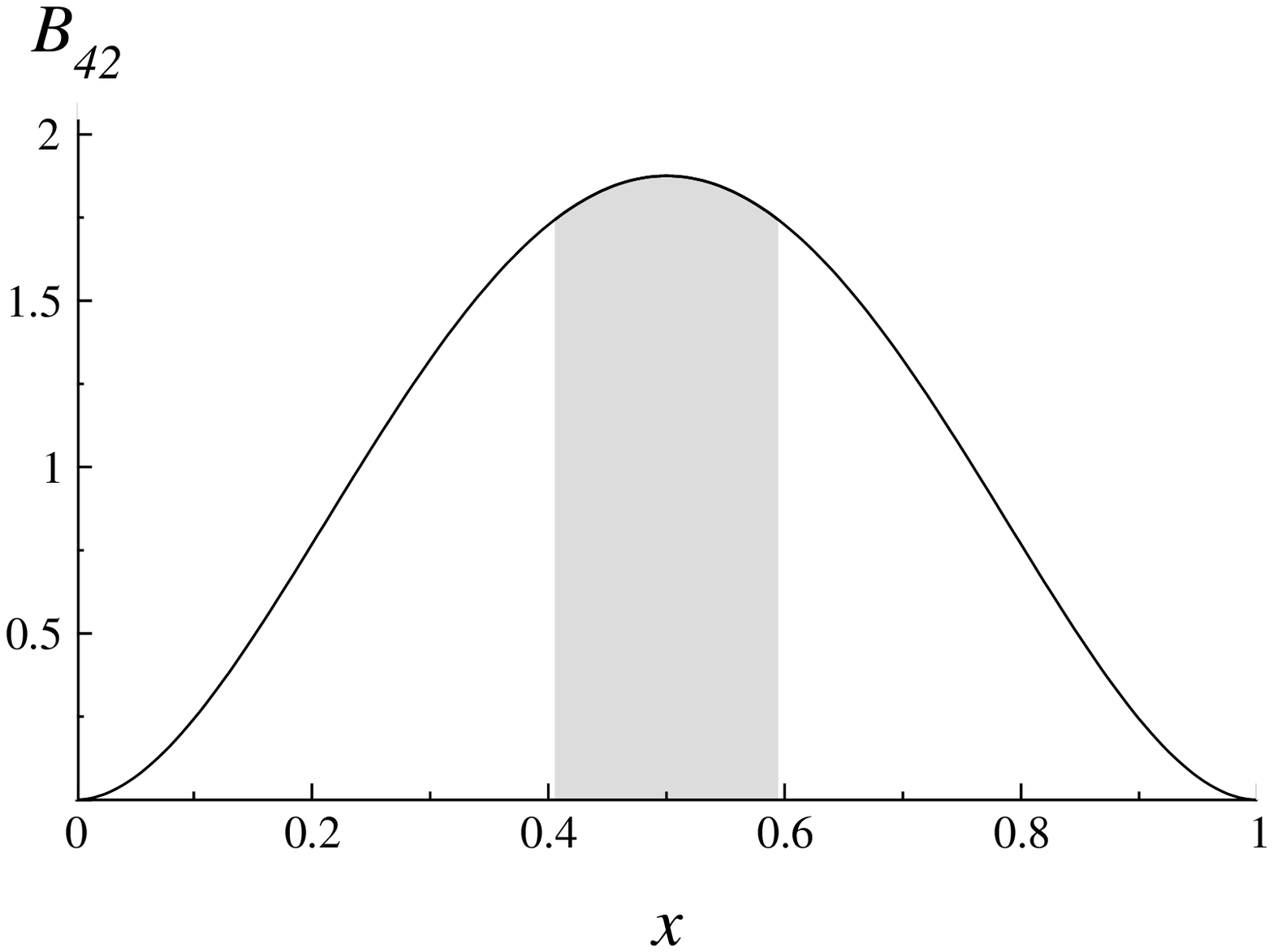}\hspace*{0.8truecm}
\includegraphics[width=0.3\textwidth]{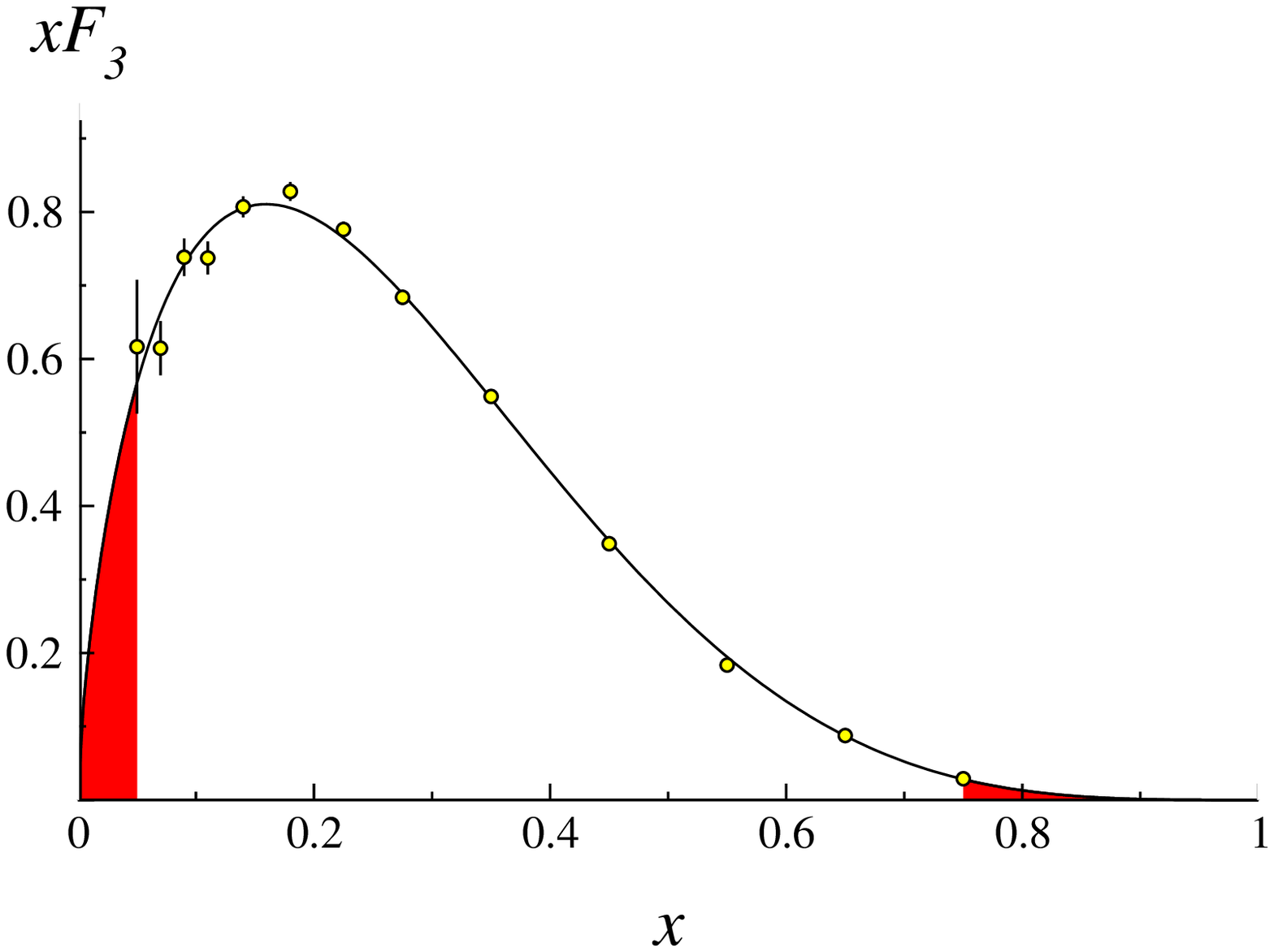}\hspace*{0.8truecm}
\includegraphics[width=0.3\textwidth]{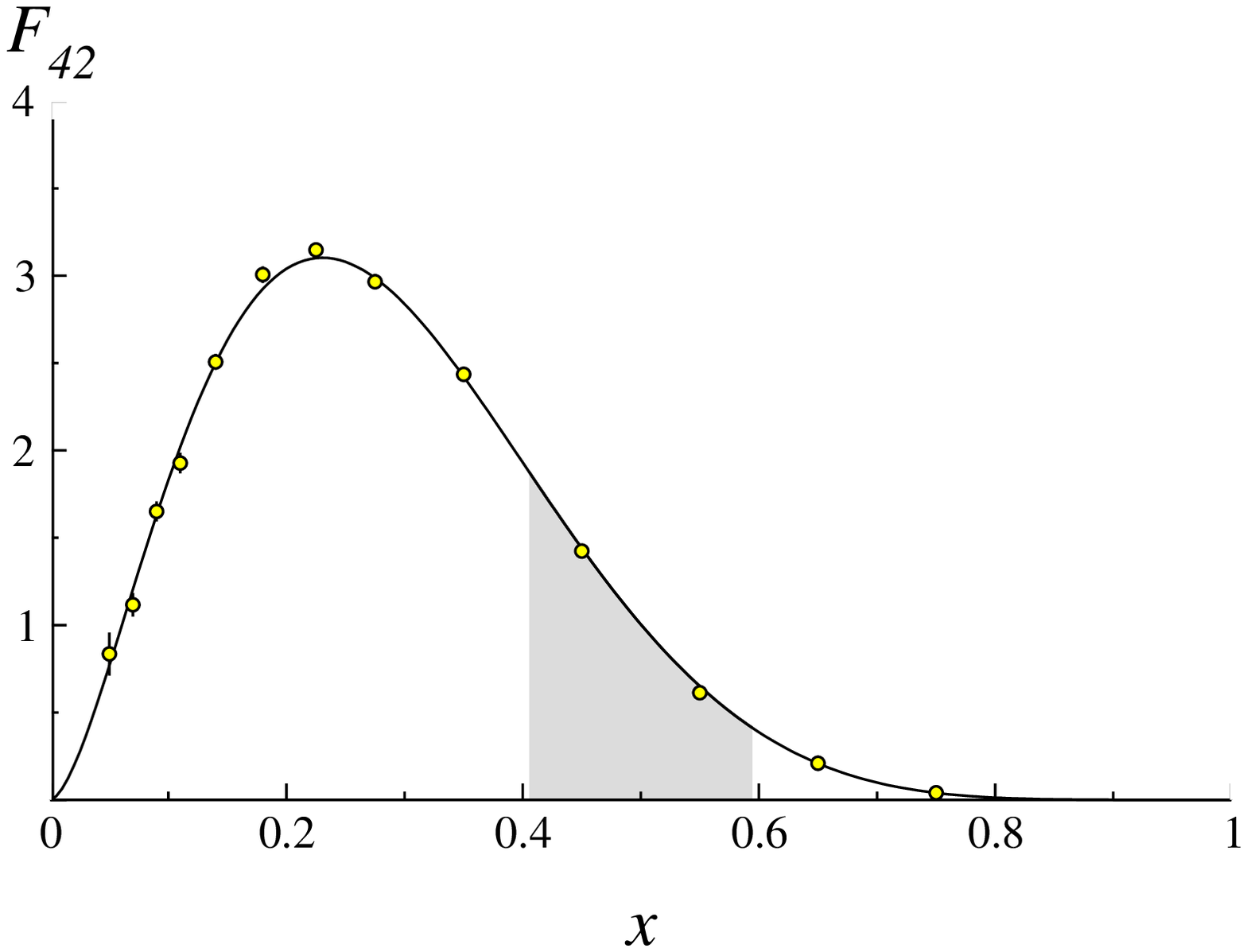}
\end{tabular}
\end{center}
\vspace*{-3truecm}
\begin{eqnarray*}
\hspace*{-0truecm}\;\times\;\frac{1}{x}\hspace*{5truecm}=\qquad
\end{eqnarray*}
\vspace*{0.8truecm}
 \caption{\textsf{Constructing the Bernstein
average, $F_{42}(Q^{2}=31.6)$. We see that the shaded (dark grey)
missing data regions almost disappear in the right hand plot. }}
\label{f:F42}
\end{figure}

The construction of an acceptable average, and the resulting
suppression of the missing data region is demonstrated in
Fig.~\ref{f:F42}. In this figure the light grey region represents
the interval
${\bar{x}}_{nk}-\frac{1}{2}\Delta{x}_{nk}{\leq}x{\leq}{\bar{x}}_{nk}+
\frac{1}{2}\Delta{x}_{nk}$ and the dark grey areas represent the
missing data regions. The small size of the  dark grey region in the
right hand plot demonstrates that this average has a negligible
dependence on the missing data regions. Note that the right hand
plot actually shows the {\it integrand} of the Bernstein average.
The average itself will be this function integrated over $[0,1]$.
\\

By expanding the integrand of Eq.~(\ref{eq:fnk1}) in powers of $x$,
we can relate the averages directly to the moments, \be
F_{nk}(Q^2)=\frac{{\Gamma(n+2)}}{\Gamma(k+1)}\sum_{l=0}^{n-k}
\frac{(-1)^l}{l!(n-k-l)!}\int_{0}^{1}x^{(k+l+1)-1}{F_3(x,Q^2)}dx\;,
\label{eq:FnkExpand} \ee and using the definition of Mellin moments
of any hadron structure function, Eq.~(\ref{eq:momxf3IV}), we have
 \be
F_{nk}(Q^2)=\frac{{\Gamma(n+2)}}{\Gamma(k+1)}\sum_{l=0}^{n-k}
\frac{(-1)^l}{l!(n-k-l)!}\;{{\cal{M}}({(k+l)+1}},Q^2)\;.
\label{eq:FnkExpandMOM}\ee

 We can only include a Bernstein average, $F_{nk}$, if we
have experimental points covering the whole range
 [${\bar{x}}_{nk}-\frac{1}{2}\Delta{x}_{nk}, {\bar{x}}_{nk}+\frac{1}{2}\Delta{x}_{nk}$]
 \cite{Max:2002},\cite{Khorramian:2004ih},\cite{SantYun}. This means that with the available experimental data we can only use
 the following 28 averages, including $F_{21}^{(exp)}(Q^2)$, $F_{31}^{(exp)}(Q^{2})$,
  $F_{42}^{(exp)}(Q^{2})$, ... .
Using Eq.~(\ref{eq:FnkExpandMOM}), the 28 Bernstein averages
${F}_{nk}({Q}^{2})$ can be written in terms of odd and even moments.
For instance: \ba
&&{F_{21}}(Q^2)=6\left({\cal{M}}(2,Q^2)-{\cal{M}}(3,Q^2)\right)\;,
\nonumber\\
&&{F_{31}}(Q^2)=24\left(0.5\;{\cal{M}}(2,Q^2)
-{\cal{M}}(3,Q^2)+0.5\;{\cal{M}}(4,Q^2)\right)\;,\nonumber\\
&&{F_{42}}(Q^2)=60\left(0.5\;{\cal{M}}(3,Q^2)
-{\cal{M}}(4,Q^2)+0.5\;{\cal{M}}(5,Q^2)\right)\;,\nonumber\\
&& \vdots \ea

\begin{figure}[tbh]
\centerline{\includegraphics[width=0.7\textwidth]{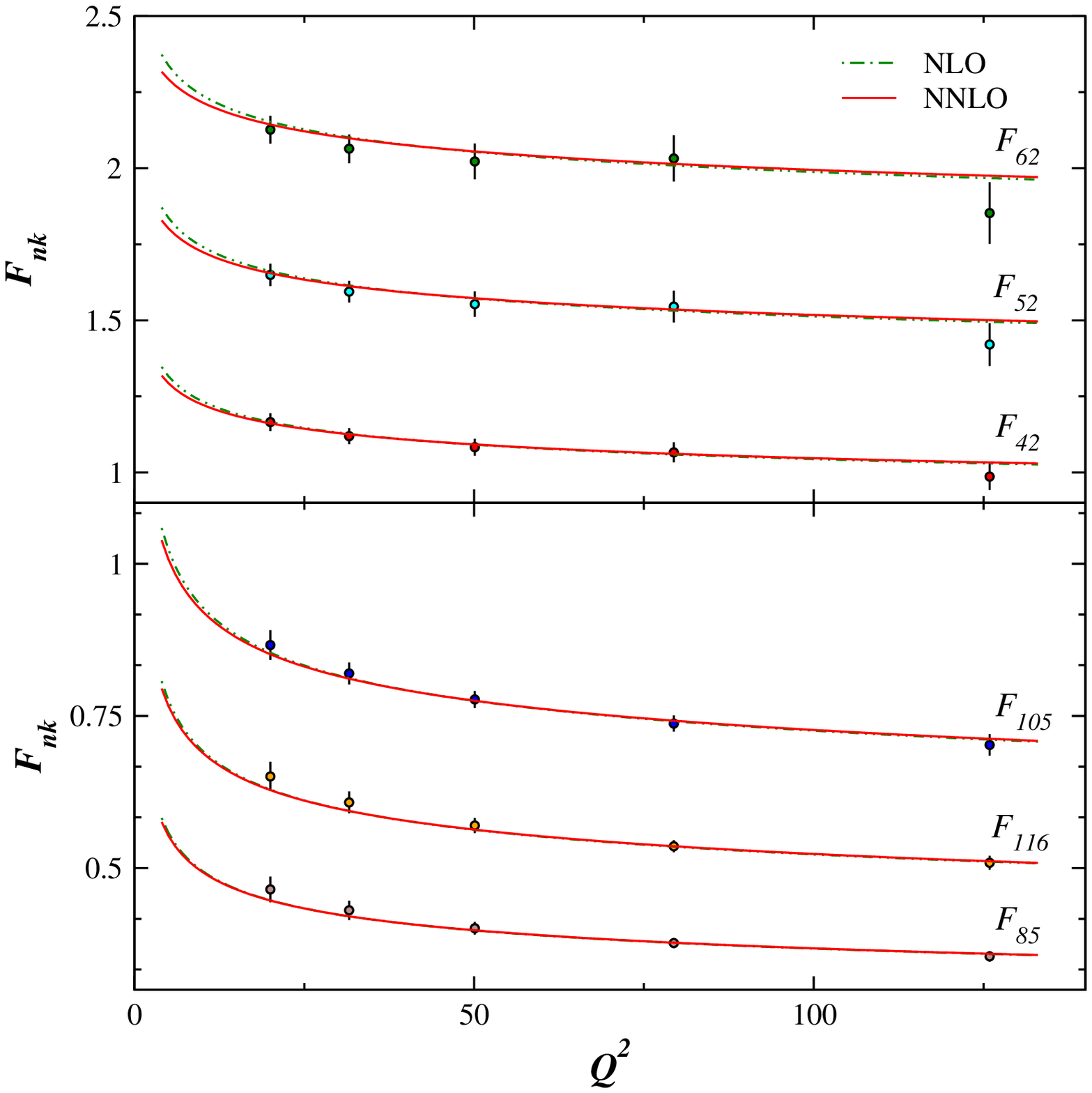}}
\caption{\textsf{NLO and NNLO fits to Bernstein averages of
$xF_3$.}} \label{pic:Fig2-FNKFIT}
\end{figure}

\begin{figure}[tbh]
\centerline{\includegraphics[width=0.8\textwidth]{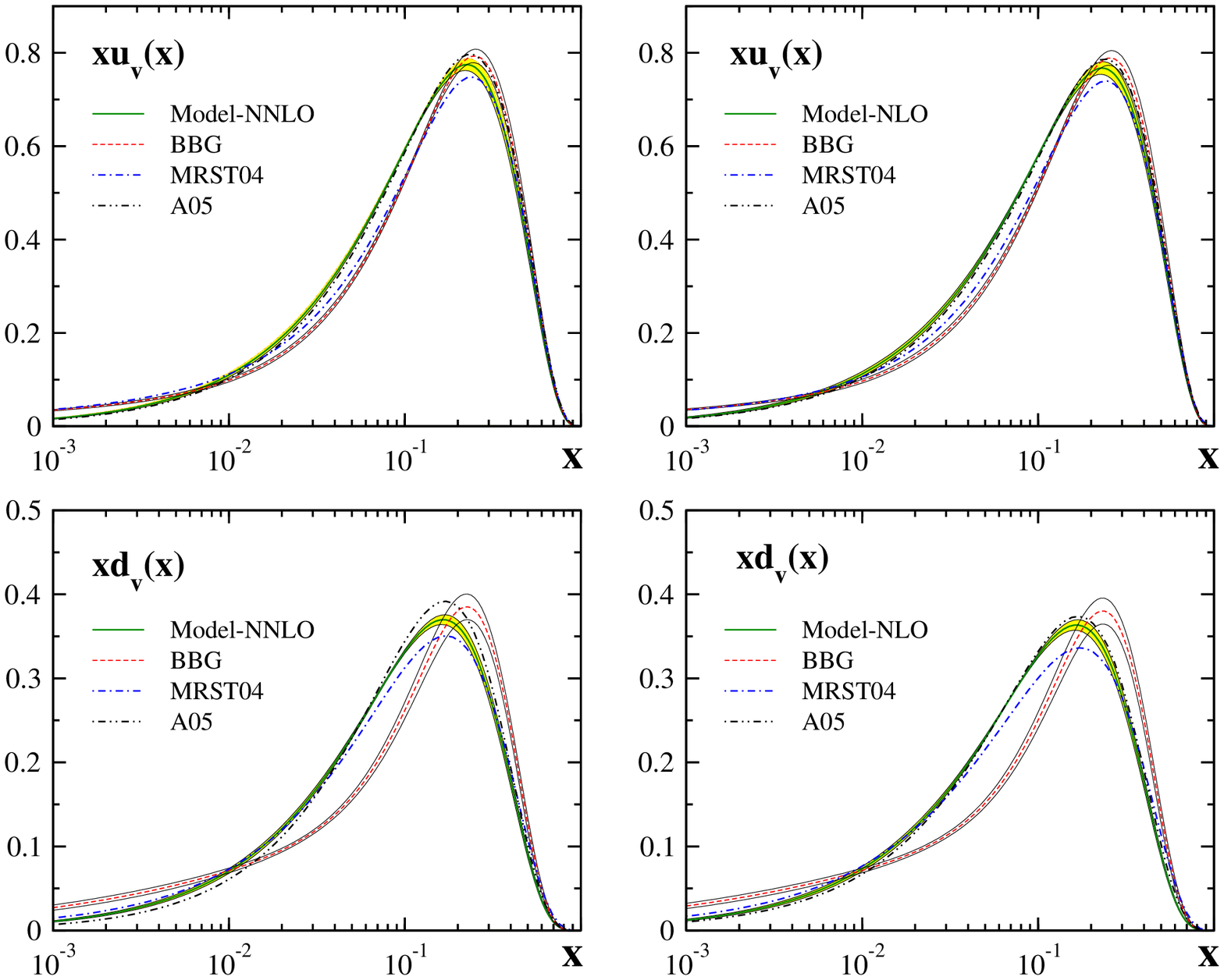}}
\caption{\textsf{The parton densities $xu_v$ and $xd_v$ at the input
scale $Q_0^2=1.0$ GeV$^2$ (solid line) compared to results obtained
from NNLO analysis (left panels) and NLO analysis (right panels) by
BBG \cite{Blumlein:BBG06} (dashed line), MRST (dashed-dotted line)
\cite{Martin:2004} and A05 (dashed-dotted-dotted line)
\cite{Alekhin:2005gq}.}} \label{pic:Fig3-partonQ0}
\end{figure}

 We can now compare the theoretical
predictions with the experimental results for the Bernstein
averages. Another restriction  we assume here, is  to ignore the
effects of moments with high order $n$ which do not strongly
constrain the fits. To obtain these experimental averages from
CCFR data \cite{CCFR:1997}, we fit $x{F_3}(x,{Q^2})$ for each bin
in ${Q}^{2}$ separately to the phenomenologically convenient
expression given in Eq.~(\ref{eq:xf3pheno}). Using
Eq.~(\ref{eq:xf3pheno}) with the fitted values of
${\cal{A}},{\cal{B}}$ and ${\cal{C}}$, one can then compute
${F}_{nk}^{(exp)}({Q}^{2})$ using Eq.~(\ref{eq:FnkExpand}), in
terms of Gamma functions. Some sample experimental Bernstein
averages are plotted in Fig.~\ref{pic:Fig2-FNKFIT} in the higher
approximations. The errors in the ${F}_{nk}^{(exp)}(Q^2)$
correspond to allowing the CCFR data for $x{F}_{3}$ to vary within
the experimental error bars, including the experimental systematic
and statistical errors \cite{CCFR:1997}.
\\
The unknown parameters according to
Eqs.~(\ref{eq:xuv},\ref{eq:xdv}) will be $a,b,c,d,e$ and $\Lambda
_{QCD}^{N_{f}}$. Thus, there are 6 parameters for each order to be
simultaneously fitted to the experimental ${F}_{nk}({Q}^{2})$
averages. Using the CERN subroutine MINUIT \cite{MINUIT:CERN}, we
defined a global ${\chi}^{2}$ for all the experimental data points
and found an acceptable fit with minimum
${\chi}^{2}/{\rm{dof}}=92.259/134=0.688$ in the LO case,
$77.452/134=0.578$ in the NLO  and $74.772/134=0.558$
 in the NNLO case. The best fit is indicated by some sample curves in
Fig.~\ref{pic:Fig2-FNKFIT}. The fitting parameters with their
uncertainties and the minimum ${\chi}^{2}$ values in each order
are listed in Table~2.
 From Eqs.(\ref{eq:xuv},\ref{eq:xdv}), we
are now able to determine the $xu_v$ and $xd_v$ at the scale of
$Q_0^2$ in higher order corrections. In
Fig.~\ref{pic:Fig3-partonQ0} we have plotted the NLO and NNLO
approximation results of $xu_v$ and $xd_v$ with correlated errors
at the input scale $Q_0^2=1.0$ GeV$^2$ (solid line) compared to
results obtained from NNLO analysis (left panels) and NLO analysis
(right panels)  by BBG \cite{Blumlein:BBG06} (dashed line), MRST
(dashed-dotted line)
\cite{Martin:2004} and A05(dashed-dotted-dotted line) \cite{Alekhin:2005gq}.\\
\begin{center}
\begin{tabular}{|c|c|c|c|c|}
\hline\hline &  & LO & NLO & NNLO \\ \hline
$u_{v}$ & $N_{u}$ & $1.952$ & $3.942$ & $5.134$ \\
& $a$ & $0.570\pm 0.016$ & $0.777\pm0.020$ & $0.830\pm0.019$ \\
& $b$ & $3.300\pm 0.089$ & $3.548\pm0.032$ & $3.724\pm0.039$ \\
& $c$ & $-0.380\pm 0.040$ & $0.410\pm0.039$ & $0.040\pm0.004$ \\
& $d$ & $4.900\pm 0.136$ & $1.500\pm0.027$ & $1.449\pm0.104$ \\
\hline
$d_{v}$ & $N_{d}$ & $1.308$ & $2.620$ & $3.348$ \\
& $e$ & $1.699\pm0.135$ & $1.563\pm0.014$ & $1.460\pm0.020$ \\
\hline \multicolumn{2}{|c|}{$\Lambda _{QCD}^{N_{f}=4},MeV$} &
$211\pm27$ & $259\pm18$ & $230\pm12$
\\ \hline\hline
\multicolumn{2}{|c|}{$\chi ^{2}/ndf$} & $92.259/134=0.688$ &
$77.452/134=0.578$ & $74.772/134=0.558$ \\ \hline\hline
\end{tabular}
 {\normalsize \\
 \textsf{\\Table~2: Parameter values of the LO, NLO,
and NNLO non-singlet QCD fit at $Q_0^2=1$ GeV$^2$.}}
\end{center}

\vspace{1 cm}
 All of the non-singlet parton distribution functions in
moment space for any order are now available, so we can use the
inverse Mellin technic to obtain the $Q^2$ evolution of valance
quark distributions which will be done in the next section.

\section{Valence quark densities in the $\bf{x}$-space}

\begin{figure}[tbh]
\centerline{\includegraphics[width=0.8\textwidth]{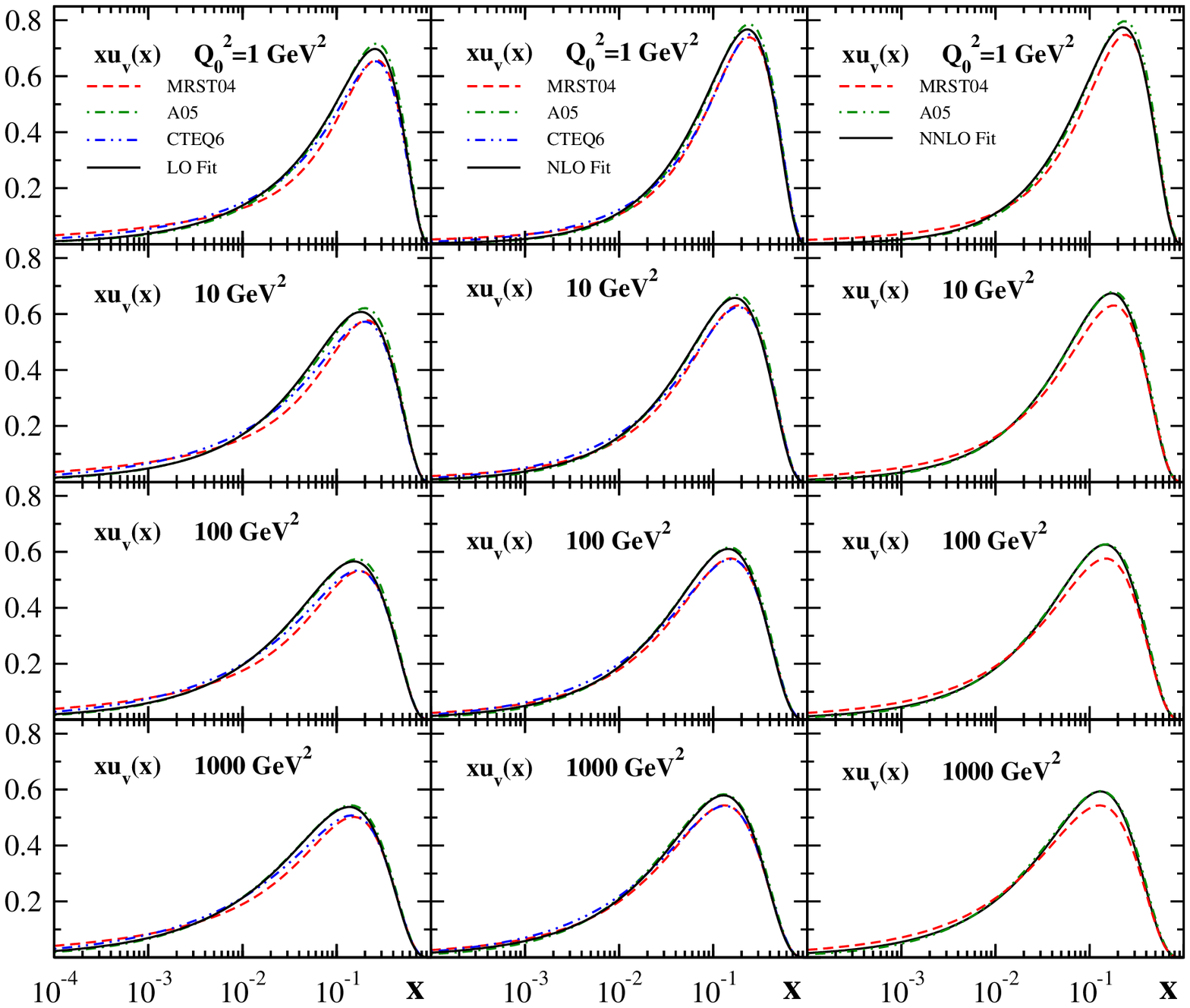}}
\caption{\textsf{The parton distribution $xu_v$ at
  some different values of $Q^{2}$. The solid line is our model, dashed line is
  the MRST model
\cite{Martin:2004}, dashed-dotted line is the A05 model
\cite{Alekhin:2005gq}
 and dashed-dotted-dotted line is the CTEQ
model \cite{Pumplin:2002vw}.}} \label{pic:Fig4-xuv}
\end{figure}

\begin{figure}[tbh]
\centerline{\includegraphics[width=0.8\textwidth]{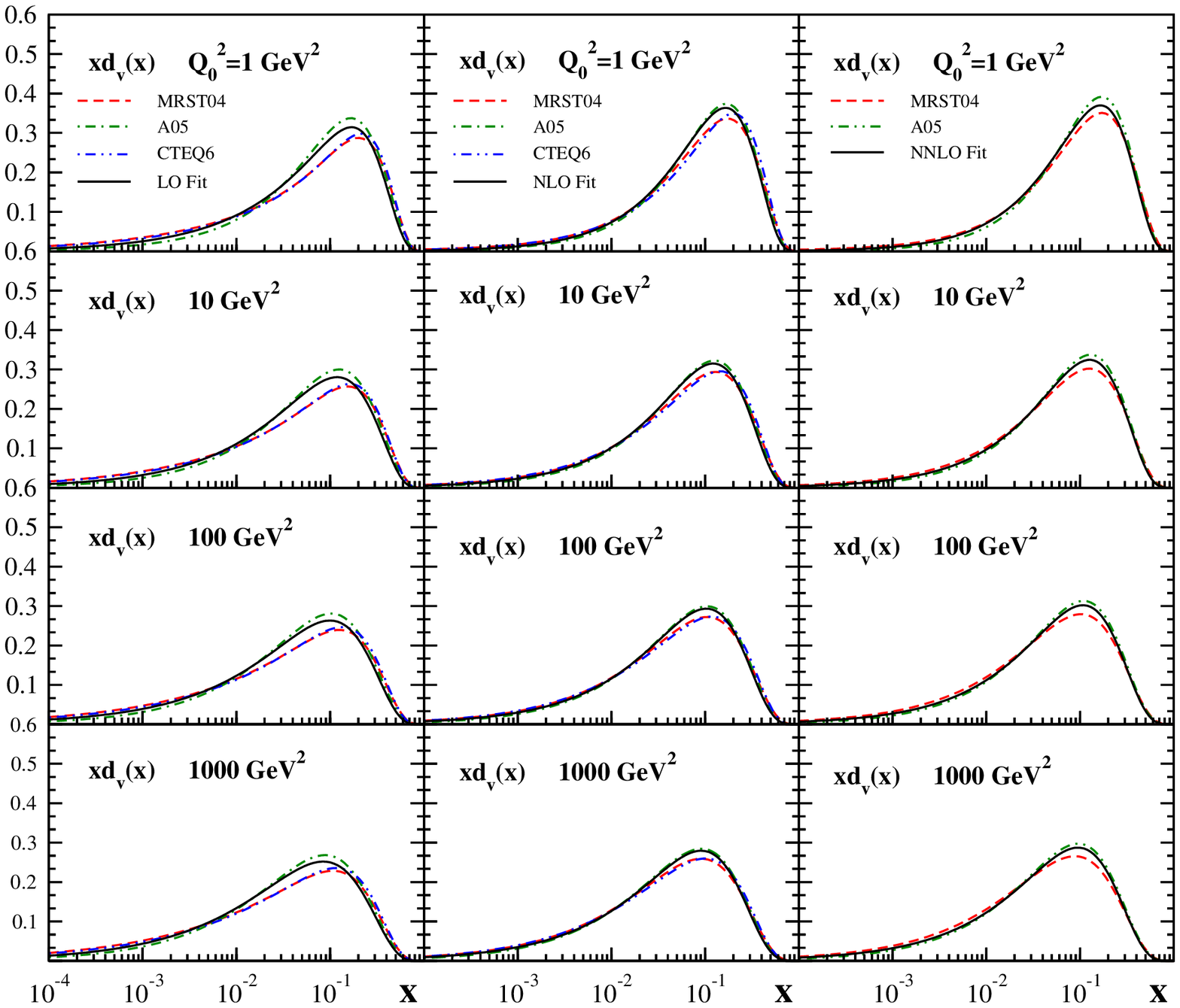}}
\caption{\textsf{The  parton distribution $xd_v$ at
  some different values of $Q^{2}$. The solid line is our model, dashed line is
  the MRST model
\cite{Martin:2004}, dashed-dotted line is the A05 model
\cite{Alekhin:2005gq}
 and dashed-dotted-dotted line is the CTEQ
model \cite{Pumplin:2002vw}.}} \label{pic:Fig5-xdv}
\end{figure}

In the previous section we parameterized the non-singlet parton
distribution functions at input scale of $Q_0^2=1$ GeV$^2$ in the
LO, NLO and NNLO approximations by using Bernstein averages method.
To obtain the non-singlet parton distribution functions in $x$-space
and for $Q^2>Q_0^2$ GeV$^2$ we need to use the non-singlet evolution
equation for parton densties to 3-loop order in
Eq.(\ref{eq:mnsNNLO}).  To obtain the $x$-dependence of parton
distributions from the $N-$%
dependent exact analytical solutions in the Mellin-moment space, one
has to perform a numerical integral in order to invert the
Mellin-transformation \cite{Graudenz:1995sk}
\begin{equation}
 q_{v}(x,Q^{2})=\frac{1}{\pi }\int_{0}^{\infty }dw Im[e^{i\phi
}x^{-c-we^{i\phi }} M_{q_v}(N=c+we^{i\phi },Q^{2})]\;,
\end{equation}
with $q_{v}=u_{v}, d_{v}$. In this equation the contour of the
integration lies on the right of all singularities of
$M_{q_v}(N=c+we^{i\phi },Q^{2})$ in the complex $N$-plane. For all
practical purposes one may choose $c\simeq 1,\phi =135^{\circ }$ and
an upper limit of integration, for any $Q^{2}$, of about $5+10/\ln
x^{-1}$, instead of infinity, which guarantees stable numerical
results \cite{GRV:90,GRV:92}. In this way, we can obtain all valence
distribution functions in fixed $Q^2$ and in $x$-space. In
Fig.~\ref{pic:Fig4-xuv} we have presented the  parton distribution
$xu_v$ at
  some different values of $Q^{2}$. These distributions were compared to LO
 , NLO and NNLO approximations with some theoretical
models \cite{Martin:2004,Alekhin:2005gq,Pumplin:2002vw}.\\
In Fig.~\ref{pic:Fig5-xdv} we have presented the same distributions
for $xd_v$. We should notice that in Figs.~\ref{pic:Fig3-partonQ0},
~\ref{pic:Fig4-xuv} and ~\ref{pic:Fig5-xdv} the minimum value of
$Q^2$ from \cite{Martin:2004}  and \cite{Pumplin:2002vw} is $1.25$
GeV$^2$
and $1.3$ GeV$^2$ respectively.\\
In Table~4 comparison of low order moments at
 $Q^2=4$ GeV$^2$ from our non-singlet NNLO QCD analysis
with the NNLO analysis BBG06~\cite{Blumlein:BBG06},
MRST04~\cite{Martin:2004}, A02~\cite{Alekhin:2005gq} and
A06~\cite{Alekhin:2006zm} has been done. \vspace{1 cm}

\begin{center}
{\normalsize
\begin{tabular}{|c|c|c|c|c|c|c|}
\hline\hline $f$ & $N$ & NNLO & BBG06 & MRST04 & A02 & A06 \\
\hline\hline
$u_{v}$ & 2 & 0.2934 $\pm$ 0.0036 & 0.2986 $\pm$ 0.0029 & 0.285 & 0.304 & 0.2947 \\
& 3 & 0.0825 $\pm$ 0.0012& 0.0871 $\pm$ 0.0011  & 0.082 & 0.087 & 0.0843 \\
& 4 & 0.0311 $\pm$ 0.0004 & 0.0333 $\pm$ 0.0005 & 0.032 & 0.033 & 0.0319 \\
\hline
$d_{v}$ & 2 & 0.1143 $\pm$ 0.0013 & 0.1239 $\pm$ 0.0026  & 0.115 & 0.120 & 0.1129 \\
& 3 & 0.0262 $\pm$ 0.0003& 0.0315 $\pm$ 0.0008 & 0.028 & 0.028 & 0.0275 \\
& 4 & 0.0083 $\pm$ 0.0001& 0.0105 $\pm$ 0.0004  & 0.009 & 0.010 & 0.0092 \\
\hline\hline
\end{tabular}

} {\textsf{\\ Table~4: Comparison of low order moments at
 $Q^2=4$ GeV$^2$ from our non-singlet NNLO QCD analysis
with the NNLO analysis BBG06~\cite{Blumlein:BBG06},
MRST04~\cite{Martin:2004}, A02~\cite{Alekhin:2005gq} and
A06~\cite{Alekhin:2006zm}.}}
\end{center}

\section{Conclusion}
\label{sec:alphaMZ}
\vspace{1mm}\noindent
 The QCD analysis is performed in LO, NLO
and NNLO based on Bernstein polynomial approach. We determine the
valence quark densities in a wide range of $x$ and $Q^2$. The QCD
scale $\Lambda_{\rm QCD}^{\rm N_f =4}$ is determined together with
the parameters of the parton distributions.  In Table~5 we have
summarized our fit results comparing $\Lambda_{\rm QCD}^{\rm N_f
=4}$ and $\alpha_s(M_Z^2)$ for  the LO, NLO and NNLO analysis. The
LO value of $\Lambda_{\rm QCD}^{\rm N_f =4}$ is found to be smaller
than the NLO value, while the NLO value comes out somewhat higher
than the NNLO value.

\begin{center}
\begin{tabular}{|c|c||c|}
\hline\hline & $\Lambda _{QCD}^{N_{f}=4}$, MeV & $\alpha
_{s}(M_Z^2)$
\\ \hline
LO & $211\pm27$ & $0.1291\pm0.0025$ \\
NLO & $259\pm18$ & $0.1150\pm0.0011$ \\
NNLO & $230\pm12$ & $0.1142\pm0.0008$ \\ \hline\hline
\end{tabular}
\\
\textsf{\\Table~5: $\Lambda_{\mathrm{QCD}}^{\mathrm{N_f=4}}$ and $%
\alpha_s(M_Z^2)$ at LO, NLO and NNLO.}
\end{center}

We compare the results of the present analysis to results
\cite{Santiago:2001mh}, \cite{Blumlein:BBG06},
\cite{Alekhin:2005gq}, [52-57] obtained in the literature at  NLO
and NNLO in Table~6, where most of the NLO values for
$\alpha_s(M_Z^2)$ presented are determined in combined singlet-
and non-singlet analysis.  The NLO values for $\alpha_s(M_Z^2)$
are larger than those at NNLO in several analysis. The difference
of both values, however, is not always the same. This is most
likely due to the type of the analysis being performed (singlet
and non-singlet, non-singlet only, etc.), in which also partly
different data sets are analyzed. Non-singlet QCD analysis were
also performed for neutrino data by using Jacobi polynomial
method. In \cite{Kataev:2001kk} the CCFR iron data on
$xF_3(x,Q^2)$ \cite{CCFR:1997} were analyzed in NLO and NNLO using
fixed moments. Likewise a NNLO analysis was performed in
\cite{Santiago:2001mh}. In \cite{Kataev:2001kk} rather large
values for $\Lambda_{\rm QCD}^{\rm N_f =4,\overline{\rm MS}}$:
$\Lambda_{\rm QCD,NLO}^{\rm N_f =4,\overline{\rm MS}} = 371 \pm 72
\MeV$, $\Lambda_{\rm QCD,NNLO}^{\rm N_f =4,\overline{\rm MS}} =
316 \pm 51 \MeV$ are obtained, which are larger than the values
obtained in the analysis based on $F_2^{p,d}(x,Q^2)$ data, still
showing the pattern that the NNLO value is lower than the NLO
value. In  Ref.~ \cite{Santiago:2001mh} one finds $\Lambda_{\rm
QCD,NLO}^{\rm N_f =4,\overline{\rm MS}}  = 281 \pm 57 \MeV$, $
\Lambda_{\rm QCD,NNLO}^{\rm N_f =4,\overline{\rm MS}} = 255 \pm 55
\MeV$.

And finally in \cite{Blumlein:BBG06} with the QCD analysis of deep
inelastic world data, the value of $\Lambda_{\rm QCD}^{\rm N_f
=4,\overline{\rm MS}} $ is reported as
\begin{eqnarray}
\Lambda_{\rm QCD,NLO}^{\rm N_f =4,\overline{\rm MS}}  &=& 265 \pm 27 \MeV\\
\Lambda_{\rm QCD,NNLO}^{\rm N_f =4,\overline{\rm MS}} &=& 226 \pm 25
\MeV~.
\end{eqnarray}
which seems close to results of the present analysis. We believe
that the difference of the reported value above, not only depends on
the type of analysis being performed (singlet and non-singlet,
non-singlet only, etc.) but also on the kind of approach ($N$-space,
x-space, etc.) have been taken.
\begin{center} {\normalsize \ }
{\normalsize
\begin{tabular}{|l|l|c|}
\hline\hline & $\alpha _{s}(M_{Z}^{2})$ & Ref. \\ \hline\hline
\textbf{NLO } &  &  \\ \hline
CTEQ6 & $0.1165 \pm 0.0065$ & \cite{Pumplin:2002vw} \\
MRST03 & $0.1165 \pm 0.0020$ & \cite{MRST03} \\
A02 & $0.1171 \pm 0.0015$ & \cite{Alekhin:2005gq} \\
ZEUS & $0.1166 \pm 0.0049$ & \cite{ZEUS_Ch} \\
H1 & $0.1150 \pm 0.0017$ & \cite{H1} \\
GRS & $0.112$ & \cite{Gluck:2006yz} \\
BBG & $0.1148 \pm 0.0019$& \cite{Blumlein:BBG06} \\ \hline Model &
$0.1150 \pm 0.0011$ &
\\ \hline\hline \textbf{NNLO} &  &  \\ \hline
MRST03 & $0.1153 \pm 0.0020$& \cite{MRST03} \\
A02 & $0.1143 \pm 0.0014$& \cite{Alekhin:2005gq} \\
SY01(ep) & $0.1166  \pm 0.0013$& \cite{Santiago:2001mh} \\
SY01($\nu $N) & $0.1153 \pm 0.0063$& \cite{Santiago:2001mh} \\
GRS & $0.111$ & \cite{Gluck:2006yz} \\
A06 & $0.1128 \pm 0.0015$& \cite{Alekhin:2006zm} \\
BBG & $0.1134^{+0.0019}_{-0.0021}$ & \cite{Blumlein:BBG06} \\ \hline Model & $0.1142 \pm 0.0008$ &  \\
\hline\hline
\end{tabular}
}
\\
{ \textsf{\\ Table~6: Comparison of $\alpha _{s}(M_{Z}^{2})$ values
from NLO, and NNLO QCD analysis.}}
\end{center}

Another important characteristic of the deep inelastic
neutrino-nucleon scattering is the Gross-Llewellyn Smith (GLS) sum
rule \cite{Gross:1969jf} \be
GLS(Q^2)=\frac{1}{2}\int_0^1\frac{xF_3^{\bar{\nu} p+\nu
p}(x,Q^2)}{x}dx\;.
 \ee

In the work of Ref.~\cite{Leung:1992yx}, the following result of the
measurement of the GLS sum at the scale $\mid Q^2\mid=3$ GeV$^2$ was
reported:

\be \label{eq:GLSCCFR} GLS(\mid Q^2\mid=3\; GeV^2)=2.5 \pm
0.018(stat.) \pm 0.078(syst.)
 \ee

In Ref.~\cite{Kataev:1994rj} the GLS($\mid Q_0^2\mid=3$ GeV$^2$)
were analyzed based on the Jacobi polynomials expansion method. The
value of GLS in the NLO approximation is reported as GLS(3
GeV$^2$)=2.446$\pm 0.081$ \cite{Kataev:1994rj}, which is in
agreement with the results Eq.~(\ref{eq:GLSCCFR}). We should notice
that in order to obtain NLO expression for the GLS sum rule one
should consider the NNLO approximation of the moments ${\cal
M}(N,Q^2)$. Following Ref.~\cite{Kataev:1994rj}, we analyze the GLS
sum rule with the corresponding perturbative QCD predictions for the
first Mellin moments, and obtain

\be \label{eq:GLSOUR} GLS(\mid Q^2\mid=3\; GeV^2)=2.40 \pm 0.06  \ee

 We hope to report on the
application of the methods employed in the present work to describe
more complicated hadron structure functions, and on using the
singlet case to extract parton densities in three loop in future
works.

\section{Acknowledgments}
We are especially grateful to G. Altarelli for fruitful suggestions,
discussions and critical remarks. We wish to thank J. Bl\"umlein for
giving us his useful and constructive comments about flavor
threshold matching. A.N.K is thanking to F. James and I. Maclaren
for discussion about MINUIT CERN program library. We would like to
thank M. M. Sheikh Jabbari, M. Ghominejad  for reading the
manuscript of this paper and for useful discussions. A. Mirjalili is
thanked for useful discussions. A.N.K is grateful to CERN for their
hospitality whilst he visited there and could amend this paper. We
acknowledge the Institute for Studies in Theoretical Physics and
Mathematics (IPM) and Semnan university for the financial support of
this project.



\end{document}